\documentclass{article}
\usepackage[utf8]{inputenc}

\usepackage[authoryear, round]{natbib}
\usepackage{times, amssymb, amsmath, amsthm}
\usepackage[hidelinks]{hyperref}

\usepackage{tensor}
\usepackage{eucal}
\usepackage{verbatim}

\usepackage{authblk}
\usepackage{braket}

\usepackage{url}

\usepackage{xargs}                      

\usepackage{graphicx}

\usepackage{xcolor}  
\usepackage[disable, colorinlistoftodos,prependcaption,textsize=tiny]{todonotes}

\newcommandx{\markb}[2][1=]{\todo[linecolor=red,backgroundcolor=red!25,bordercolor=red,#1]{#2}}
\newcommandx{\chris}[2][1=]{\todo[linecolor=blue,backgroundcolor=blue!25,bordercolor=blue,#1]{#2}}
\newcommandx{\niels}[2][1=]{\todo[linecolor=green,backgroundcolor=green,bordercolor=green,#1]{#2}}

\usepackage{tikz}
\tikzset{
  treenode/.style = {shape=rectangle, rounded corners,
                     draw, align=center,
                     top color=white, bottom color=blue!20},
  root/.style     = {treenode, font=\Large, bottom color=red!30},
  env/.style      = {treenode, font=\ttfamily\normalsize},
  dummy/.style    = {circle,draw}
}

\definecolor{cblue}{RGB}{100,5,255}    
\definecolor{cred}{RGB}{255,10,10} 
\definecolor{cgreen}{RGB}{5,165,20}  
\definecolor{corange}{rgb}{1.0,0.49,0.0}

\begin{document}

\title{Noether's first theorem and the energy-momentum tensor ambiguity problem\thanks{To appear in: The Physics and Philosophy of Noether's Theorems; James Read, Bryan Roberts and Nicholas Teh (Eds.); Cambridge University Press, forthcoming.}}

\author[1,2]{Mark Robert Baker\thanks{mrbaker@stfx.ca}}
\author[1,3]{Niels Linnemann\thanks{niels.linnemann@uni-bremen.de}}
\author[1,4]{Chris Smeenk\thanks{csmeenk2@uwo.ca}}

\affil[1]{Rotman Institute of Philosophy, University of Western Ontario, 1151 Richmond Street North, London, Ontario, CA-N6A5B7}

\affil[2]{Department of Physics, St. Francis Xavier University, 4130 University Ave, Antigonish, NS, CA-B2G2W5}

\affil[3]{Institut für Philosophie, Universität Bremen, Postfach 330 440, Enrique-Schmidt-Str. 7, D-28334 Bremen}
 
\affil[4]{Department of Philosophy, University of Western Ontario, 1151 Richmond Street North, London, Ontario, CA-N6A5B7}

\date{\today}

\maketitle

\begin{quote}
\textit{Dedicated to the late Bessel-Hagen, who when alive had his habilitation thesis thrown into the sea, and even now must feel as if his work was lying somewhere on the seabed.}
\end{quote}

\abstract{
Noether's theorems are widely praised as some of the most beautiful and useful results in physics. However, if one reads the majority of standard texts and literature on the application of Noether's first theorem to field theory, one immediately finds that the ``canonical Noether energy-momentum tensor" derived from the 4-parameter translation of the Poincar\'e group does not correspond to what's widely accepted as the ``physical'' energy-momentum tensor for central theories such as electrodynamics. This gives the impression that Noether's first theorem is in some sense not working. In recognition of this issue, common practice is to ``improve" the canonical Noether energy-momentum tensor by adding suitable ad-hoc ``improvement" terms that will convert the canonical expression into the desired result. On the other hand, a less common but distinct method developed by Bessel-Hagen considers gauge symmetries as well as coordinate symmetries when applying Noether's first theorem; this allows one to uniquely derive the accepted physical energy-momentum tensor without the need for any ad-hoc improvement terms in theories with exactly gauge invariant actions. Given these two distinct methods to obtain an energy-momentum tensor, the question arises as to whether one of these methods corresponds to a preferable application of Noether's first theorem.  Using the converse of Noether's first theorem, we show that the Bessel-Hagen type transformations are uniquely selected in the case of electrodynamics, which powerfully dissolves the methodological ambiguity at hand. We then go on to consider how this line of argument applies to a variety of other cases, including in particular the challenge of defining an energy-momentum tensor for the gravitational field in linearized gravity. Finally, we put the search for proper Noether energy-momentum tensors into context with recent claims that Noether's theorem and its converse make statements on equivalence classes of symmetries and conservation laws: We aim to identify clearly the limitations of this latter move, and develop our position by contrast with recent philosophical discussions about how symmetries relate to the representational capacities of our theories.
}

\tableofcontents

\section{Introduction}

Physicists have long exploited symmetries to simplify problems. In Lagrangian mechanics, cyclic coordinates (that is, generalized coordinates $q_i$ such that $\partial \mathcal{L} / \partial q_i = 0$ for the Lagrangian $\mathcal{L}$) signal the presence of a symmetry, and the Euler-Lagrange equations imply that the associated conjugate momenta $p_i$ are conserved.\footnote{See \cite{butterfield2006symmetry} for a pedagogical presentation.} It is hard to understate the practical importance of finding conserved quantities, thereby reducing the number of variables and making it much easier to find solutions. Noether's celebrated (1918) paper significantly clarified the mathematical structure underlying earlier results.\footnote{\citet{noether1918}, translated into English by \cite{kosmann2011}.}  The rich line of work stemming from her seminal contribution has elucidated three intertwined aspects of physical theories: laws, symmetries, and conservation principles.  

Conventional wisdom now holds that Noether's first theorem and its converse universally link a certain kind of continuous symmetry (such as Poincaré translation) to a certain kind of conserved current (such as the energy-momentum tensor).\footnote{The metaphysical work of \cite{Lange2007}, for instance, rests on this.} Although based on a kernel of truth, this conventional wisdom reflects an overly simplified picture of the mathematical physics. As a starting point for the discussion below, consider the following specific claim often taken to follow from the Noether machinery: a subset of the variational symmetries of the action, namely spatial and temporal translations, are associated with energy-momentum conservation. Here we encounter an immediate difficulty: applying Noether's first theorem in the context of field theory (as described in \S \ref{section:NoetherEM}), the canonical variational symmetry and 4-parameter translation subgroup of the Poincar\'e group yields what is called the ``canonical Noether energy-momentum tensor'' ($T^{\mu\nu}_C$). For most classical field theories, the canonical tensor lacks features required for a physically sensible energy-momentum tensor, and differs from known physical energy-momentum tensors established in other ways.\footnote{We will focus on energy-momentum in the ensuing discussion, but similar issues arise for other conserved currents, such as angular momentum.} Such results raise two related challenges to the conventional wisdom: do the quantities that actually follow from applying Noether's theorem have a clear physical interpretation, and does Noether's theorem need to be supplemented in order to derive physically meaningful conserved quantities? Particularly striking is the existence of inequivalent definitions of the energy-momentum tensor, a central physical quantity in any classical field theory.

Typical textbook presentations leave the impression that Noether's theorem fails to yield the correct energy-momentum tensor. They mention the unappealing features of $T^{\mu\nu}_C$, and then immediately propose a fix. Such fixes amount to variations on a theme going back to \cite{belinfante1940current}, who added the divergence of a so-called superpotential to $T^{\mu\nu}_C$ such that a new ``Belinfante" energy-momentum tensor $T^{\mu\nu}_B = T^{\mu\nu}_C + \partial_\alpha b^{[\mu\alpha]\nu}$ recovers the correct answer for electrodynamics\footnote{Unless otherwise indicated, the ensuing discussion focuses on sourceless electrodynamics for ease of exposition.} if an on-shell condition is imposed (discussed in more detail below in \S \ref{section:TwoMethods}). This does not follow directly from Noether's theorem itself, suggesting that some form of ``improvement'' is needed to find physically meaningful conserved quantities (as we discuss in \S \ref{section:methodI}). But any such improvement approach has the unsavoury air of devising a series of poorly justified steps to arrive at an answer found in the back of the book. What happens when we do not already know, or have independent ways of finding, the correct form for the energy-momentum tensor?

Thankfully there is another approach, albeit much less common in the literature. Noether's first theorem is often taken to be the result of a general identity applied to a specific type of action symmetry, namely finite symmetries such as the Poincar\'e group. But we can also consider the consequences of this identity for finite \emph{and} infinite (e.g. gauge) symmetries of the action --- although the first theorem is usually assumed to apply only to finite symmetry groups, Noether herself suggested a broader formulation of the theorem.\footnote{The end of Noether's statement of theorem I holds that it also applies to an infinite-parameter group: ``Dieser Satz gilt auch noch im Grenzfall von unendlich vielen Parametern." (``This proposition [Satz] also holds in the limit of infinitely many parameters.") To our mind it is ambiguous whether this clarification is to be taken as part of the theorem or as an additional theorem; it is in this sense of Noether's first theorem, however, that  Poincar\'e translation symmetry \textit{and} gauge symmetries of the action can work together. Our main interest is in what Noether's mathematical framework allows one to prove based on symmetries of the action, following from the general identity below (\ref{noetherfirstelectrodynamics}); whether one regards a particular result as falling within the scope of the ``first theorem'' is partly a terminological issue. Thanks to Harvey Brown for pressing us on this point.}

As we will see below, in the case of electrodynamics this leads directly to the correct energy-momentum tensor. This was Bessel-Hagen's neglected contribution, inspired in part by Noether herself \citep{bessel1921erhaltungssatze}. By contrast with the ``improvement'' approach, we will argue that this approach exploits all of the relevant variational symmetries of the action in applying Noether's theorem.  We will argue for the superiority of this approach based on using the converse of Noether's first theorem, which correctly identifies the proper variational symmetries (those derived using Method II in \S \ref{SectionMethodII}) of the Lagrangian from the accepted form of the energy-momentum tensor. The derivation of the canonical stress-energy tensor fails to use the full power of the mathematical machinery that Noether has given us by considering only a restricted subset of the variational symmetries. Thus, at least in the context of Lagrangian field theories in flat spacetime, the conventional wisdom of a universal linkage between symmetries and conservation laws can be refined to that of a linkage between a specific variational symmetry --- to be introduced below --- and the set of physical conservation laws for the theory.

This line of argument does not address Lagrangian field theories in curved spacetime, which lack the global symmetries needed to obtain the energy-momentum tensor via Noether's theorem.\footnote{See also \cite{holman2010}, who argues for generalizations of Noether's theorems in Minkowski spacetime from considerations of diffeormorphism invariance, and based on these he claims to show how to obtain the correct stress-energy tensor.} Physicists then typically use Hilbert's definition of the energy-momentum tensor $T^{\gamma\rho}_H = \frac{2}{\sqrt{-g}} \frac{\delta \mathcal{L}}{\delta  g_{\gamma\rho}}$, which is sometimes referred to as the metric energy-momentum tensor. The Hilbert energy-momentum tensor is by definition symmetric, thereby avoiding one of the major flaws of the canonical Noether tensor. From this expression, the Hilbert energy-momentum tensor in Minkowski spacetime is defined as the curved spacetime Hilbert energy-momentum tensor with all metric tensors $g_{\mu\nu}$ replaced by the Minkowski metric $\eta_{\mu\nu}$. \cite{blaschke2016} outlines the distinction between the curved spacetime and Minkowski spacetime definitions of the Hilbert tensor.\footnote{ The Hilbert energy-momentum tensor in Minkowski spacetime ($T^{\gamma\rho}_{H,\eta} = \frac{2}{\sqrt{-g}} \frac{\delta \mathcal{L}}{\delta  g_{\gamma\rho}} \big|_{g = \eta}$) is not, in general, equivalent to energy-momentum tensors derived via Noether's theorem from the 4-parameter Poincaré translation, see \cite{baker2021a}.}   In any case, our discussion will focus on the status of the energy-momentum tensor derived from Noether's first theorem in flat (Minkowski) spacetime.

The plan for the paper is as follows.  This section concludes with a tribute to Bessel-Hagen.  We then briefly introduce the energy-momentum tensor, and review properties required for it to be regarded as physically reasonable.  In the next two sections, we use classical electrodynamics as a simple case to introduce Noether's first theorem and its subtleties. Section \ref{section:NoetherEM} introduces Noether's first theorem. Section \ref{section:TwoMethods} considers the two approaches described above for defining Noether currents, focusing on the energy-momentum tensor: Method I uses translation symmetries to yield the canonical energy-momentum tensor, which requires ``improvement'' terms to yield the correct energy-momentum tensor; Method II, by contrast, considers a broader class of variational symmetries and leads directly to the correct energy-momentum tensor. We then argue in favour of the second method based on the converse of Noether's theorem. Section \ref{ApplicationsSection} considers how this line of argument applies to a variety of other cases, including in particular the challenge of defining an energy-momentum tensor for the gravitational field in linearized gravity. Section \ref{section:Equivalence} brings out one of the themes running through the discussion, namely the challenge of tracking the physical significance of these structural properties of field theories.  There are several steps in the early sections where it is tempting to describe both symmetries and conserved currents only up to an equivalence class. We aim to identify clearly the limitations of this move, and develop our position by contrast with recent philosophical discussions about how symmetries relate to the representational capacities of our theories (considering, in particular, Brown's contribution in this volume). Finally we discuss the outlook and conclusions of our work in section \ref{section:Conclusions}.

\subsection{An ode to Bessel-Hagen}

Alongside these systematic aims, we want to use the occasion to clarify the contribution of Erich Bessel-Hagen (1898 - 1946) to the Noether machinery. On the one hand, Bessel-Hagen seems to be often wrongly treated only as an originator of the generalisation of Noether's theorem to invariance under symmetry transformations of the action \textit{up to a boundary term}.
\niels{Brown's point 6: In footnote 7, the paper by Brown and Holland (2004) is cited as an example of the wrongful sole attribution to Bessel-Hagen of the mentioned generalisation. But on p. 1134 of this paper and in footnote 10 therein it is noted that Noether anticipated this generalisation. See also footnote 37 in Brown (in this volume). NL: There must have been some misunderstanding upon rewriting. What was intended to be said here, is that people only recognize about Bessel-Hagen that he has managed to provide a generalisation of Noether to divergences --- but not that he is the originator of how to use Noether's first theory in the context of gauge theories. MB: Agreed with Niels NL: OK DONE WITH THIS (see also reply file)}\footnote{Importantly, it is not correct to see Bessel-Hagen as the \textit{sole} originator of this generalisation in any case: as \cite{kosmann2011} (see her section 4.2) notes, Bessel-Hagen himself (ambiguously) acknowledges his debt to Noether herself (to a certain degree, at least):

\begin{quote}
Zuerst gebe ich die beiden E. Noetherschen Sätze an, und zwar in einer etwas allgemeineren Fassung als sie in der zitierten Note stehen. Ich verdanke diese einer mündlichen Mitteilung von Fräulein Emmy Noether selbst. (\cite{bessel1921erhaltungssatze}, p. 260) (I first present the two Noetherian propositions, albeit in a slightly more general fashion than they can be found in the cited note. I owe these propositions to an oral communication by Miss Emmy Noether herself. (Own translation))
\end{quote}}$^{\text{, }}$\footnote{We do not apply this generalization in our article. See \S \ref{ApplicationsSection} and \S \ref{section:Equivalence} for more discussion.}

\noindent On the other hand, Bessel-Hagen does not seem to be widely known for his central contribution in that very same paper, namely the introduction of what we call Method II: the application of Noether's first theorem in light of gauge symmetries when deriving the complete set of conformal conservation laws for classical electrodynamics. Bessel-Hagen's work has been independently reproduced --- by, among others, \cite{eriksen1980,montesinos2006}; the reception of his paper in the English speaking world, however, suffered from the fact that a translation first appeared in 2006 \citep{besseltranslation2006}, arguably much too late. Even though the method has resurfaced in some textbooks as well \citep{burgess2002,scheck2012}, it remains relatively unknown in the wider physics literature. The Bessel-Hagen method has recently been applied to a wide class of special relativistic field theories \citep{baker2021c}.  Here we contrast the more common textbook approach with Bessel-Hagen's method, primarily focusing on the simple case of electrodynamics.

\section{Energy-Momentum Tensors}
\label{section:EMT}

Einstein took the general formulation of conservation laws in terms of the energy-momentum tensor to be ``the most important new advance in the theory of relativity'' (as of 1912).  The energy-momentum tensor has a central role in the new conception of mechanics and field theory, as Einstein went on to emphasize:\footnote{See \cite{janssen2006classical} for an insightful discussion of the importance of the energy-momentum tensor in the transition to relativistic mechanics.}  
\begin{quote}
    To every kind of material process we want to study, we have to assign a symmetric tensor $T_{\mu\nu}$ [...]   The problem to be solved always consists in finding out how $T_{\mu\nu}$ is to be formed from the variables characterizing the processes under consideration. If several processes can be isolated in the energy-momentum balance that take place in the same region, we have to assign to each individual process its own stress-energy tensor ($T_{\mu\nu}^1$, and etc.) and set $T_{\mu\nu}$ equal to the sum of these individual tensors. (CPAE Vol. 4, Doc. 1, [p. 63])
\end{quote}
Strikingly, Einstein treats all ``material processes,'' whether they involve electromagnetic fields or matter as described by continuum mechanics, as on a par:  the fundamental dynamical quantity in each case is the energy-momentum tensor.  How then are we to find an appropriate $T_{\mu\nu}$ for various processes we aim to describe?

 Before turning to that question, recall that the energy-momentum tensor (also known as the stress-energy tensor) encodes information regarding energy-momentum densities and fluxes for different kinds of ``material processes.''  In relativistic mechanics this is all captured in a single rank-two tensor, $T^{\mu\nu}$:  the $T^{00}$ component represents energy density, the $T^{0i}$ and $T^{i0}$ components represent energy and momentum flux, respectively, and the $T^{ij}$ components represent stresses (where $i,j = 1,2,3$).

As an illustration, the energy-momentum tensor for electromagnetism provides a compact summary of familiar facts about the electromagnetic field. Minkowski formulated electromagnetism in terms of the field strength tensor $F_{\mu\nu} = \partial_{\mu}A_{\nu} - \partial_{\nu}A_{\mu}$, where $A_{\mu}$ is the vector potential.  The energy-momentum tensor takes the following form:\footnote{The energy density of electromagnetic fields is given by $U = \frac{1}{2}(\epsilon_0 E^2 + \frac{1}{\mu_0} B^2)$, the Poynting vector $\vec{S} = \frac{1}{\mu_0} \vec{E} \times \vec{B} $ represents energy flux, and the Maxwell stress tensor $\sigma_{ij}$ represents stress and momentum fluxes. We can express the energy-momentum tensor in terms of these quantities as follows:  $T^{\mu\nu} = 
\begin{pmatrix}
   - U  & - S_x/ c & - S_y/ c   &  - S_z/ c 
\\
- S_x/ c   &    \sigma_{xx} & \sigma_{xy} & \sigma_{xz} 
\\
  - S_y/ c   &    \sigma_{yx} & \sigma_{yy} & \sigma_{yz}
\\
 - S_z/ c   &     \sigma_{zx} & \sigma_{zy} & \sigma_{zz}
 \end{pmatrix} $.} 
\begin{equation}
T^{\mu\nu} = F^{\mu\alpha} F^\nu_{\ \alpha} - \frac{1}{4} \eta^{\mu\nu} F_{\alpha\beta} F^{\alpha\beta}.  \label{electrodynamicsemt}
\end{equation}
\noindent The invariance of the field strength tensor $F^{\mu\nu}$ under gauge transformations ($A'_{\mu} = A_{\mu} + \partial_{\mu} \phi$, for a scalar gauge parameter $\phi$) implies gauge invariance of $T^{\mu\nu}$.  More generally, we require gauge invariance of $T^{\mu\nu}$ because the energy-momentum tensor represents observable quantities directly. In this case, we have constructed the energy-momentum tensor based on what we already know about the relevant field.  Historically, von Laue extended this constructive approach, writing down appropriate energy-momentum tensors for extended stressed bodies, relativistic fluids, and other cases, based on prior knowledge about energy and momentum in each case. 

To what extent can we determine the form of $T^{\mu\nu}$ for a new classical field $\psi$ (whether scalar, vector, tensor,...) based on general principles, or on specific features of $\psi$'s dynamics?  There are two main types of constraints a tensor would be expected to satisfy to be plausibly interpreted as representing energy-momentum of the field.  The first set of constraints stem from the idea that all matter fields ``carry positive energy-momentum.'' More formally, for arbitrary regions of spacetime $R$, $T^{\mu\nu}$ vanishes on $R$ iff the field $\psi$ vanishes.\footnote{Here we are setting aside fields with negative energy density, for which $T^{\mu\nu}$ could vanish through a cancellation of positive and negative energy densities.  This condition has to be formulated with greater care for quantum fields, which necessarily admit negative expectation values for the energy density at spacetime points, but versions of this condition have been proposed for open regions.} Further constraints can be imposed to capture the idea that the energy-momentum is positive, and that energy-momentum flows respect the causal structure of relativistic spacetime. One fundamental requirement of this kind is that the energy density (the $T^{00}$ component) is bounded from below, so that the field cannot serve as an infinite energy source. Gravity is sensitive to the absolute value of the energy, so that it is meaningful to differentiate positive and negative energies for fields coupled to gravity. Further constraints can then be imposed: the weak energy condition, for example, requires that the energy density (the $T^{00}$ component) is non-negative, as measured by all observers.  The dominant energy condition holds if, in addition, momentum fluxes stay within the light cone.  There is a long list of other energy conditions that have been used to prove results such as the singularity theorems.\footnote{See \cite{curiel2017primer} for a comprehensive review of energy conditions, their status, and their role in sundry theorems.}

A second set of constraints, and the main focus of the ensuing discussion, regards symmetries and conserved quantities.  We will take the satisfaction of an appropriate conservation principle as a defining feature of energy-momentum.\footnote{This leaves open the possibility that there are fields, such as the metric field in general relativity, that lack an energy-momentum tensor in this sense.} Given an appropriate $T^{\mu\nu}$, the on-shell conservation principle can be succinctly stated: $\partial_{\mu}T^{\mu\nu} = k^{\nu}$. (For a free field, $k^{\nu}=0$, otherwise $k^{\nu}$ represents an external force density.\footnote{In the case of electrodynamics sourced by $J^\alpha$ this is the force density $f^\nu = \partial_\mu T^{\mu \nu} = F^\nu_{\ \alpha} J^\alpha$ which includes the Lorentz force density in the spatial components ($f^i = \partial_\mu T^{\mu i} = J^{\rho} F^{i}_{\ \rho} = \rho \vec{E} + \vec{J} \times \vec{B}$).}) In classical mechanics, the conservation of energy and momentum stem from space-time translation symmetries, so it is plausible to begin by constructing a tensor combining the conserved currents associated with these symmetries. The variational symmetries of an action $S$ consist of the transformations that leave $S$ invariant,\footnote{There are cases where certain transformations leave $S$ invariant only up to a boundary term; see \S \ref{ApplicationsSection} and \S \ref{section:Equivalence} for more discussion.}\niels{Brown's point 7: On p. 6, in the penultimate paragraph, variational symmetries are defined in terms of transformations leaving the action invariant (as Noether did in her 1918 paper). What happened to the (Noether-Bessel-Hagen) generalisation to invariance up to a surface term? The derivative in the equation in this line is a total derivative. The same is true for the derivatives in equation (2), other than derivatives of the As. Shouldn’t these details be mentioned? NL: this is a good point I think on which as a group still have a disagreement? please let us discuss MB: We don't consider up to surface term in this article, just mention that this is an additional feature of Bessel-Hagen's paper... we should discuss this point not sure I agreed with Brown here} and Noether's first theorem associates a conserved quantity with each element of the finite group of transformations. In electrodynamics these symmetry constraints, the variational symmetries of the action for coordinates (conformal symmetries) and fields (gauge symmetries), are what is required to obtain the known conservation laws, which we detail in \S \ref{SectionMethodII}; the 15 conservation laws are associated to the (finite) 15 parameter conformal group of transformations.

Other properties of the energy-momentum tensor follow from symmetries of the field theory for specific types of fields. For classical field theories with conformal symmetry, for example, the energy-momentum tensor will be trace-free so that the conformal $C^{\rho\alpha}$ and dilatation $D^\rho$ tensors are conserved.\footnote{In the case of electrodynamics, this statement is directly related to the associated quantum particles are massless (see \cite{Garg}, p. 563).}

We take these two types of constraints as requirements that a rank two tensor must satisfy to be a plausible candidate\niels{reviewr: I understand that that authors are proposing necessary not sufficient conditions for physicality. But if the authors are not themselves making a commitment to which specific version of their criteria are to be satisfied, then I worry that `appropriate’ is doing too much heavy lifting in both constraints. One can easily gerrymander an energy condition and a derivative operator for an `unphysical’ S-E tensor rending it `physical’ by the authors’ lights. Since the whole point of this exercise is to rule out certain S-E tensors as physical, even if they do not present us with sufficient conditions, I feel like the reader would benefit from the authors suggesting how their criteria could rule out obviously `bad’ stress energy tensors as unphysical. NL: maybe being Noetherian is a sufficient criterion} for an ``energy-momentum tensor'' of particular importance in considering energy-momentum tensors proposed for a new field $\psi$ rather than constructed based on prior knowledge. Perusing the physics literature suggests that these two types of constraints do not suffice to determine a unique choice:  there are several proposed, apparently inequivalent, candidates for the ``energy-momentum tensor for $\psi$'' (for a variety of different fields). Our overall aim below is to argue against this view.  Several of the candidate energy-momentum tensors may not be worthy of the name. Take, for example, the question of whether we should require that the energy-momentum is symmetric under exchange of indices ($T^{\mu\nu} = T^{\nu\mu}$). Failure of this to hold in a mechanical system would lead to torque and the possibility of unlimited angular acceleration.\footnote{An angular momentum tensor $M^{\rho\mu\nu} = x^{\mu}T^{\rho\nu} - x^{\nu}T^{\rho\mu}$ is conserved iff $T^{\mu\nu}$ is symmetric. The angular momentum relative to a given event chosen as an origin can be obtained by integrating $M^{\rho\mu\nu}$; see Chapter 5 of \cite{MTW} for further discussion of this point, and regarding the general properties of energy-momentum tensors.} Similar problems arise in field theories. Just as in the case of mechanics, an anti-symmetric energy-momentum tensor would entail failure of angular momentum conservation if the dynamical fields are all scalar quantities.  This does not hold for the more general case of tensorial fields, however, which can have non-trivial spin degrees of freedom that also contribute to the total angular momentum (as we will see below in more detail).  

Although this case is more subtle than the criteria emphasized above, a preference for symmetric rank-two tensors already rules out many candidates discussed in the literature. We will further argue that we need to take into account more than just spacetime translations in building the energy-momentum tensor out of conserved currents, as we will illustrate next by considering the case of electromagnetism in more detail.

\section{Noether's first theorem for classical electrodynamics}
\label{section:NoetherEM}

Noether's first theorem, applied to a particular Lagrangian density, yields a relationship between the Euler-Lagrange expressions and Noether current of the theory of the form $E^A \delta \phi_A + \partial_\mu J^{\mu} = 0$, where $E^A$ is the Euler-Lagrange expression for the rank-$A$ field and $J^\mu$ is the Noether current. This relationship is derived by substituting a particular Lagrangian density and action symmetries into a general identity which is sometimes referred to as the Noether identity.\footnote{The Noether identity is the starting point for both of Noether's first and second theorems. It is sometimes referred to as the Noether off-shell condition.} For the Lagrangian density of electrodynamics $\mathcal{L} = - \frac{1}{4} F_{\mu \nu} F^{\mu \nu}$, the Noether identity is \citep[e.g.,][]{fomin1963}:

\begin{equation}
\left( \frac{\partial \mathcal{L}}{\partial A_\nu}
- \partial_\rho \frac{\partial \mathcal{L}}{\partial (\partial_\rho A_\nu)} \right) \bar{\delta A_\nu} 
+ \partial_\rho \left( 
\eta^{\rho\beta} \mathcal{L} \delta x_\beta
+ \frac{\partial \mathcal{L}}{\partial (\partial_\rho A_\nu)}\bar{\delta A_\nu} 
  \right)
 = 0, \label{noetherfirstelectrodynamics}
\end{equation}

\noindent
where

\begin{equation}
\bar{\delta A_\nu} = - \partial^\beta A_\nu \delta x_\beta + \delta A_\nu \label{gentran}
\end{equation}

\noindent
is the complete set of symmetry transformations that are directly obtained upon deriving the Noether identity (\ref{noetherfirstelectrodynamics}).\footnote{Equation (\ref{noetherfirstelectrodynamics}) corresponds to Equation 12 in Noether's paper \citep{kosmann2011} for the specific Lagrangian of classical electrodynamics, and Equation (\ref{gentran}) corresponds to Equation 9 in Noether's paper. Our $x_\beta$ (coordinates) correspond to her independent variables $x_n$ and our $A_\nu$ (fields) correspond to her functions of these independent variables $u_i$.}

The two methods we will discuss diverge with regard to the general form of the transformations $\bar{\delta A_\nu}$.
Consider first the difference between $\bar{\delta A_\nu}$ and $\delta A_\nu$:  the non-bar transformation of fields is the difference in transformed fields as a function of their respective coordinates,

\begin{equation}
\delta A_\nu = A'_\nu(x') - A_\nu(x). \label{nonbartrans}
\end{equation}

\noindent
By contrast, the bar transformations of fields is the difference in transformed fields as a function of the same (non-transformed) coordinates,

\begin{equation}
\bar{\delta A_\nu} = A'_\nu(x) - A_\nu(x)  \label{bartrans},
\end{equation}

\noindent
where the bar notation is adopted from Noether's paper for this particular transformation in her Equation 9. The following subsections treat the three types of transformations ($\bar{\delta A_\nu}$): (1) the two terms which correspond to the Lie derivative of the fields (canonical and contragredient transformations) and (2) the term which corresponds to gauge symmetries of the action.

\subsection{Transformations associated to the Lie derivative}

In this subsection, we describe two of the contributions to the transformations of fields $\bar{\delta A_\nu}$ arising from infinitesimal change of coordinates $\delta x^\nu$. These two transformations follow directly from the Lie derivative of the four potential $A_\nu$ with respect to the infinitesimal change in coordinates $\delta x$, 

\begin{equation}
\pounds_{{}_{\delta x}} A_\nu = - \delta x^\beta \partial_\beta A_\nu  -  A_\beta \partial_\nu \delta x^{\beta} = \delta_C A_\nu + \delta_T A_\nu.
\end{equation}

\noindent The Lie derivative represents the coordinate invariant change of a tensor field along the flow of a vector field, which is in this case the infinitesimal change in coordinates $\delta x$.

We have denoted the two terms in this expression as $\delta_C A_\nu$ and $\delta_T A_\nu$, respectively. The first term, $\delta_C A_\nu = - \delta x^\beta \partial_\beta A_\nu$, is exactly what is found in the first term of (\ref{gentran}). This term alone is used to derive the canonical Noether energy-momentum tensor when $\delta x^\beta = a^\beta$ is the 4-parameter Poincar\'e translation, thus we will refer to this as the canonical transformations. The second term, $\delta_T A_\nu =  -  A_\beta \partial_\nu \delta x^{\beta}$, we will refer to as contragredient transformations as Bessel-Hagen did in his article; they are associated to the transformation properties of a tensor. This contribution is zero for $\delta x^\beta = a^\beta$, and thus does not factor into energy-momentum tensor discussion. However, for any non-constant $\delta x^\beta$ this contribution is nonzero and essential for deriving the associated conserved tensors, such as the angular momentum tensor resulting from the remaining parameters of the Poincar\'e group.

\subsubsection{Canonical transformations and the canonical Noether energy-momentum tensor}

If we restrict ourselves to canonical transformations, $- \partial^\beta A_\nu \delta x_\beta$  with $\delta A_\nu = 0$, and no gauge symmetries, we have only

\begin{equation}
\bar{\delta A_\nu} = \delta_C A_\nu = - \partial^\beta A_\nu \delta x_\beta \label{canonicaltrans}
\end{equation}

\noindent to substitute into the Noether identity (\ref{noetherfirstelectrodynamics}). We will use $\delta_C A_\nu$ to indicate these canonical transformations. We then obtain:
\begin{equation}
\left( \frac{\partial \mathcal{L}}{\partial A_\nu}
- \partial_\rho \frac{\partial \mathcal{L}}{\partial (\partial_\rho A_\nu)} \right) \bar{\delta A_\nu} 
+ \partial_\rho \left( 
[\eta^{\rho\beta} \mathcal{L} 
- \frac{\partial \mathcal{L}}{\partial (\partial_\rho A_\nu)} \partial^\beta A_\nu] \delta x_\beta
  \right)
 = 0 
\end{equation}

\noindent The square brackets contain what is known as the ``canonical Noether energy-momentum tensor" $T^{\rho\beta}_C$ for a Lagrangian density of the form $\partial A \partial A$ such as $\mathcal{L} = - \frac{1}{4} F_{\mu \nu} F^{\mu \nu}$ of electrodynamics. In the case of the 4-parameter Poincar\'e translation $\delta x_\beta = a_\beta$, we can factor out the constant $a_\beta$ from the divergence yielding $E^\nu \bar{\delta A_\nu} + a_\beta \partial_\rho T^{\rho\beta}_C = 0$, where

\begin{equation}
T^{\rho\beta}_C = \eta^{\rho\beta} \mathcal{L} 
- \frac{\partial \mathcal{L}}{\partial (\partial_\rho A_\nu)} \partial^\beta A_\nu \label{gencanonemt}
\end{equation}

\noindent and $E^\nu$ is the Euler-Lagrange expression (in the case of $\mathcal{L} = - \frac{1}{4} F_{\mu \nu} F^{\mu \nu}$, $E^\nu$ are the non-homogeneous Maxwell's equations). 

However, substituting the Lagrangian density for electrodynamics $\mathcal{L} = - \frac{1}{4} F_{\mu \nu} F^{\mu \nu}$, where $F_{\alpha\beta} = \partial_\alpha A_\beta - \partial_\beta A_\alpha$ and $\frac{\partial \mathcal{L}}{\partial (\partial_\rho A_\nu)} = - F^{\rho\nu}$, yields

\begin{equation}
T^{\rho\beta}_C = F^{\rho\nu} \partial^\beta A_\nu 
- \frac{1}{4} \eta^{\rho\beta} F_{\mu \nu} F^{\mu \nu} \label{electrodynamicscanonical},
\end{equation}

\noindent the canonical Noether energy-momentum tensor for classical electrodynamics. By contrast, the accepted energy-momentum tensor $T^{\mu\nu}$ for the theory is given by eqn (\ref{electrodynamicsemt}) above.  

This specific case illustrates two distinct problems for the canonical energy-momentum tensor that hold more broadly. First, the result simply does not match with an independently motivated expression for the energy-momentum tensor, based on an understanding of energy and momentum densities and fluxes for the relevant field. Second, the canonical tensor lacks essential properties: in general it is neither symmetric, nor gauge invariant, nor trace-free. There are special cases where some of these properties hold. For example, a symmetric tensor follows from (\ref{gencanonemt}) for a Klein-Gordon scalar field; yet even then, there are alternative tensors which improve on the canonical expression by being trace-free \citep{callan1970}.  In some of these cases, it may not be obvious whether the canonical tensor or some other candidate tensor is to be preferred. We do not claim to have a way to resolve this debate across the board; rather, there are several clear cases (like electromagnetism) where the canonical energy-momentum tensor fails to have the right form.

\subsubsection{Contragredient transformations}

The non-bar transformation of fields $\delta A_\nu$ (second term in (\ref{gentran})) is referred to by Bessel-Hagen as being associated to the ``contragredient" transformations of the fields, which in current treatments follow simply from the definition of a contravariant tensor (in this case vector):
\begin{equation}
A'^\nu(x') = \frac{\partial x'^\nu}{\partial x^{\mu}} A^\mu(x)
\end{equation}

\noindent
Inserting this into the contravariant form of (\ref{nonbartrans}) we have for $\delta A^\nu$,

\begin{equation}
\delta A^\nu = \frac{\partial x'^\nu}{\partial x^{\mu}} A^\mu(x) - A^\nu(x). \label{contra1}
\end{equation}

\noindent
If we consider the transformation of coordinates,

\begin{equation}
x'^{\nu} = x^{\nu} + \delta x^{\nu}  \label{coordtrans}
\end{equation}

\noindent
In particular, then, $\frac{\partial x'^\nu}{\partial x^{\mu}} = \delta^{\nu}_{\mu} + \partial_{\mu} (\delta x^{\nu})$. Substituting (\ref{coordtrans}) into (\ref{contra1}) we have $\delta A^\nu = A^\mu \partial_\mu \delta x^\nu$. To determine the covariant form of this expression, we can consider the identity $A^\mu A_\mu = A'^\mu A'_\mu$ as a function of their respective coordinates, and solve for the transformation $\delta_T A_\nu$, which is exactly the contragredient transformation presented by Bessel-Hagen in his equation 18,

\begin{equation}
\delta_T A_\nu = -  A_\mu \partial_\nu \delta x^{\mu} \label{contra2}
\end{equation}

\noindent
where $\delta_T A_\nu$ indicates that this is the transformation based on the definition of a tensor $T$. Note that we require the covariant form of this transformation due to our presentation of the Noether identity in (\ref{noetherfirstelectrodynamics}).

For higher rank tensors this contribution can easily become quite complicated. However, in the case of energy-momentum tensor derivation, when we have the 4-parameter Poincar\'e translation $\delta x^{\mu} = a^\mu$ --- regardless of Method I or Method II --- $\delta_T A_\nu = 0$ since $a^\mu$ is a constant. For this reason, since most of the discussions of conservation laws focus solely on the energy-momentum tensor at the expense of other conserved quantities such as angular momentum, this contribution usually drops out of the picture. (Yet we need the contragredient transformations for the derivation of conservation laws linked to non-constant coordinate symmetries $\delta x^{\mu}$.)

\subsection{Gauge (field) symmetries of the action}

There is also the possibility of gauge (field) symmetries of the action, often overlooked from the perspective of Noether's first theorem because they are thought to be relevant only to Noether's second theorem. The Bessel-Hagen et al. approach uses these symmetries as well to derive the known conservation laws of electrodynamics directly from Noether's general identity. 

Mixing the complete set of coordinate and field symmetries is essential to obtaining the accepted energy-momentum tensor of electrodynamics. To highlight this point we briefly touch on Noether's second theorem, again with a focus on electrodynamics.

Noether's second theorem is another application of the Noether identity in (\ref{noetherfirstelectrodynamics}). The basic idea behind Noether's second theorem is that field (gauge) symmetries that leave the action invariant (in her case, the ``infinite continuous group" of transformations of the functions $u$) can be integrated by parts to remove the derivatives on the field transformations; neglecting boundary terms (instead of keeping them as in the case of the Noether current in the first theorem) results in an identity in terms of the Euler-Lagrange expressions.\footnote{See Equation 16 in \cite{noether1918} and associated discussion for statements on Noether's second theorem. Notably, the converse also holds, namely that the existence of such identity implies invariance of the action under an infinite continuous group.}

\noindent

In the case of electrodynamics, discarding boundary terms (all terms under a total divergence) leaves the standard Euler-Lagrange expression, Maxwell's $\partial_\rho F^{\rho\nu}$, as,

\begin{equation}
\partial_\rho F^{\rho\nu} \bar{\delta A_\nu} 
 = 0 \label{ELclassic}
\end{equation}

\noindent
Now taking the gauge transformation $A'_\nu = A_\nu + \partial_\nu \phi$ (where $\phi$ is a scalar), we have,

\begin{equation}
\bar{\delta A_\nu} = \delta_g A_\nu = \partial_\nu \phi \label{gaugetrans}
\end{equation}

\noindent
where we denote $\delta_g A_\nu$ to emphasize the transformation associated to the gauge symmetry of the action. From (\ref{ELclassic}) and (\ref{gaugetrans}) we therefore have $\partial_\rho F^{\rho\nu} \partial_\nu \phi
 = 0$. Integrating by parts and discarding the resulting boundary term we are left with the well known identity for Noether's second theorem in electrodynamics,
 
 \begin{equation}
\partial_\rho \partial_\nu F^{\rho\nu} \phi = 0
\end{equation}

\noindent
and thus $\partial_\rho \partial_\nu F^{\rho\nu} = 0$. It is the incorporation of this transformation (\ref{gaugetrans}) that is then also essential for directly deriving the complete set of conservation laws from Noether's first theorem, including the accepted energy-momentum tensor (\ref{electrodynamicsemt}). By use of the converse of Noether's second theorem we have a concrete methodology for obtaining the variational gauge symmetry $\bar{\delta A_\nu}$ that is required for application of Method II.

\subsection{Summary}
\label{SummaryNoetherQuote}

In summary, Noether's identity (\ref{noetherfirstelectrodynamics}) can be used to obtain a relationship between the Euler-Lagrange expressions and conservation laws for field theories such as electrodynamics . 

The Noether current depends on the coordinate symmetry transformation $\delta x_\beta$ and field symmetry transformations $\bar{\delta A_\nu}$; i.e. any symmetry transformation of the action must be introduced through these contributions in order to derive corresponding on-shell conserved currents. We distinguished three main types of transformations of fields, which can be simultaneously applied to (\ref{noetherfirstelectrodynamics}) in the form,

 \begin{equation}
\bar{\delta A_\nu} = \delta_C A_\nu + \delta_T A_\nu + \delta_g A_\nu \label{alltrans}
\end{equation}

\noindent
where we have the canonical transformations (\ref{canonicaltrans}), contragredient transformations (\ref{contra2}) and gauge transformations (\ref{gaugetrans}). In the case of electrodynamics, this gives

 \begin{equation}
\bar{\delta A_\nu} = - \partial^\mu A_\nu \delta x_\mu -  A_\mu \partial_\nu \delta x^{\mu} + \partial_\nu \phi \label{alltrans2}
\end{equation}

These transformations are the complete set required to derive conservation laws in standard field theories such as electrodynamics. Bessel-Hagen derived all 15 conservation laws of electrodynamics which are associated to the 15 parameter conformal group of infinitesimal coordinate transformations,

 \begin{equation}
\delta x_\alpha = a_\alpha + \omega_{\alpha\beta} x^\beta + S x_\alpha + 2 \xi_\nu x_\alpha x^\nu  - \xi_\alpha x_\nu x^\nu \label{conformaltrans}
\end{equation}

\par\noindent In the case of the four-parameter Poincar\'e translation $\delta x_\beta = a_\beta$, the coordinate symmetry associated to energy-momentum tensor derivation, we have $\delta_T A_\nu = 0$ leaving only two contributions to the transformation of fields $\bar{\delta A_\nu}$.\footnote{The term $\omega_{\alpha\beta} x^\beta$, associated to the angular momentum tensor $M^{\rho\alpha\beta} = x^\alpha T^{\rho\beta} -  x^\beta T^{\rho\alpha}$, consists of the remaining 6 parameters in the Poincar\'e group through the antisymmetric parameter $\omega_{\alpha\beta}$. Terms $S x_\alpha$ and $2 \xi_\nu x_\alpha x^\nu  - \xi_\alpha x_\nu x^\nu$ correspond to the dilatation tensor $D^\rho$ and conformal tensor $C^{\rho\alpha}$. Direct substitution of the various terms in (\ref{conformaltrans}) into (\ref{alltrans2}) give the $\bar{\delta A_\nu}$ which can be directly substituted into (\ref{noetherfirstelectrodynamics}) to derive known physical conservation laws in e.g. electrodynamics. Transformations (\ref{conformaltrans}) can be found by solving the conformal Killing's equation. See \cite{rosen1972} for a self-contained derivation of the conformal invariance of electrodynamics.}

Notably, the presence of mixed coordinate and field transformations should be no surprise to anyone who has actually read Noether's paper, as she explicitly admits that her first theorem holds for a combination of the symmetries:

\label{mixedgroup}

\begin{quote}
In the case of a ``mixed group,'' if one assumes similarly that $\Delta x$ and $\Delta u$ are linear in the $\varepsilon$ and the $p(x)$, one sees that, by setting the $p(x)$ and the $\varepsilon$ successively equal to zero, divergence relations ... as well as identities .. are satisfied. (translated from \cite{noether1918}, p. 243)\footnote{``Setzt man entsprechend einer ,gemischten Gruppe' $\Delta x$ und $\Delta u$ linear in den $\varepsilon$ und den $p(x)$ an, so sieht man, indem man einmal die $p(x)$, einmal die $\varepsilon$ Null setzt, daß sowohl Divergenzrelationen ..., wie Abhängigkeiten ... bestehen."}
\end{quote}

\noindent
where the equation 13 she refers to is the Noether identity associated to her first theorem that we consider in this article. It is exactly this freedom that Bessel-Hagen, in fact in consultation with Noether herself, used to apply the first theorem successfully to electrodynamics; this is the topic of the section on Method II.

In the section on Method I, we will discuss the case when $\delta_g A_\nu = 0$, i.e. when we only use the canonical transformations, yielding the canonical Noether energy-momentum tensor $T^{\mu\nu}_C$. Since this is not the correct energy-momentum tensor $T^{\mu\nu}$ of electrodynamics, various ad-hoc ``improvements'' have been considered in the literature that add terms to $T^{\mu\nu}_C$ in order to obtain the desired result. In the section on Method II, we will discuss the less common method in the literature, which does not make restrictions on $\bar{\delta A_\nu}$ and keeps the most general (\ref{alltrans}). In this case, the well known $T^{\mu\nu}$ of electrodynamics is directly derived with no ad-hoc ``improvements'' needed.

\section{A Tale of Two Methods}

\label{section:TwoMethods}

\subsection{Method I: Canonical Tensor plus `Improvements'}

\label{section:methodI}

In this section, we consider the most common method in the literature for deriving the generally accepted energy-momentum tensor `from' Noether's first theorem. As emphasized above, the canonical Noether tensor $T^{\mu\nu}_C$ obtained on this approach (\ref{electrodynamicscanonical}) differs from the accepted energy-momentum tensor (\ref{electrodynamicsemt}). Hence Noether's theorem apparently fails to properly identify the conserved quantities associated with symmetries even in the most familiar case. On our view, this ``canonical'' result reflects a basic mistake:  it does not take into account all of the relevant variational symmetries needed to build an energy-momentum tensor.  We will see how to employ Noether's theorem more effectively to do so through what we call Method II in the next section (\S \ref{SectionMethodII}).

Usually the the canonical energy-momentum tensor is ``improved'' by adding specific terms, such as the divergence of a superpotential and terms proportional to the equations of motion.\footnote{There are a number of proposals regarding how to ``improve'' the energy-momentum tensor in the literature; see \citet{forger2004,blaschke2016} for recent surveys.} We give an example of this for electrodynamics in this section.  One could put the task --- a bit provocatively --- as follows: Given that $T^{\rho\beta}_C$ is not the result we wanted (or expected), what terms can we add to get the correct answer? Of course this is an ad-hoc approach to fixing the problem, but if it is the best available method we have to obtain the accepted $T^{\rho\beta}$, one might just bite the bullet.\footnote{It is worth noting that proponents of Method I are usually unaware of the Bessel-Hagen et. al approach.}

The required ``improvement" term in the case of electrodynamics is simply the difference between (\ref{electrodynamicsemt}) and the canonical expression (\ref{electrodynamicscanonical}),

\begin{equation}
\label{DifferenceT}
T^{\rho\beta} - T^{\rho\beta}_C = - F^{\rho\nu} \partial_\nu A^\beta.
\end{equation}

\noindent
All of the various improvements for electrodynamics in the literature ultimately need to give us this term on the right hand side, at minimum after imposing on-shell conditions (\cite{forger2004,blaschke2016}). The challenge is how to get the correct tensor, by starting from the canonical expression, and adding ``improvement'' terms through a well-defined procedure. We will briefly discuss the Belinfante improvement procedure since it is by far the most commonly adopted in the literature.\footnote{\cite{belinfante1940current} is commonly cited as the origin of the Belinfante improvement procedure, also known as the Belinfante symmetrization procedure, but it is not clear to what extent he shared the motivations of later work regarding the ``improvement'' of energy-momentum tensors..}

But before doing so, it is worth discussing the general idea of improvement by superpotentials and terms proportional to the equations of motion; together these form the bulk of possible ``improvement'' terms. Superpotentials have the form $\Psi^{[\rho\alpha]\sigma}$, the divergence of which $\partial_{\alpha} \Psi^{[\rho\alpha]\sigma}$ can be added to an energy-momentum tensor without affecting on-shell conservation. This is because indices $[\rho\alpha]$ are anti-symmetric; the divergence of the divergence of a superpotential $\partial_\rho \partial_{\alpha} \Psi^{[\rho\alpha]\sigma}$ is identically zero off-shell. Adding a superpotential to a Noether energy-momentum tensor for a specific Lagrangian does not spoil conservation (the superpotential is conserved as a mathematical identity on its own), yet doing so may lead to an energy-momentum tensor with the required properties. Terms that vanish on-shell, i.e. terms proportional to the equations of motion, can also be added while preserving on-shell equivalence; in practice terms of this type often must also be added to obtain the accepted form of the energy-momentum tensor.

For electrodynamics, the difference between the accepted and canonical energy-momentum tensors is given by (\ref{DifferenceT}) above.  Writing the extra term as the divergence of a superpotential, we have $- F^{\rho\nu} \partial_\nu A^\beta = \partial_\alpha [- F^{\rho\alpha} A^\beta] +  A^\beta  \partial_\nu F^{\rho\nu}$, where $\Psi^{[\rho\alpha]\sigma} = - F^{\rho\alpha} A^\beta$, and Maxwell's equations $E^\rho = \partial_\nu F^{\nu\rho}$. Thus, we have,

\begin{equation}
T^{\rho\beta} = T^{\rho\beta}_C + \partial_\alpha \Psi^{[\rho\alpha]\beta} - A^\beta  E^\rho  \label{backsuper}
\end{equation}

\noindent The Belinfante improvement procedure yields exactly the same superpotential; the divergence of this superpotential as well as a term proportional to the equation of motion can be used to recover the accepted energy-momentum tensor in cases such as electrodynamics. Therefore just by knowing (\ref{electrodynamicsemt}) we know the form of the required additional terms. The Belinfante procedure provides a derivation of this superpotential which we will detail in the following subsection.

\subsubsection{The Belinfante symmetrization procedure}

Turning an arbitrary tensor into a symmetric tensor is in principle straightforward: decompose the tensor into a symmetric and antisymmetric part, and then add a new contribution to cancel out the antisymmetric part (in this case, from a superpotential). But more interestingly, arguably, \cite{belinfante1940current}\footnote{Through the help of an uncited Dr. Podolansky (presumably theoretical physicist Dr. Julius Podolanski), see \cite{belinfante1940current}.} showed that a suitable superpotential of this kind can be derived --- and he argued that it is related to the spin angular-momentum of the model.\footnote{Since the ``total'' angular momentum tensor $M^{\rho\alpha\beta} = x^\alpha T^{\rho\beta} -  x^\beta T^{\rho\alpha}$ is based on the symmetric energy-momentum tensor $T^{\mu\nu}$, decomposing it into parts (including ``spin'' angular momentum), one can obtain terms missing in the canonical Noether tensor $T^{\mu\nu}_C$.  The missing terms have the appropriate form for spin angular momentum; if this tensor vanishes the canonical tensor is symmetric, otherwise we have to follow the procedure described in the text.} 
We wish to add the divergence of the Belinfante superpotential $ \partial_\alpha b^{[\rho\alpha]\sigma}$ to the canonical expression (\ref{electrodynamicscanonical}) to form the Belinfante tensor $T^{\rho\sigma}_B$,\footnote{We note that the Belinfante tensor $T^{\rho\sigma}_B$ is sometimes referred to as the Belinfante-Rosenfeld tensor, since \cite{rosenfeld1940} independently came to some of these results and published them shortly after \cite{belinfante1939} first presented them. 

}

\begin{equation}
T^{\rho\sigma}_B = T^{\rho\sigma}_C + \partial_\alpha b^{[\rho\alpha]\sigma}
\end{equation}

\noindent
where the superpotential $b^{[\rho\gamma]\sigma}$ is defined by a combination of the spin angular momentum tensor $S^{\rho [\sigma \gamma]}$ of the form (\cite{belinfante1940current}):

\begin{equation}
b^{[\rho\gamma]\sigma} =
\frac{1}{2} (- S^{\rho [\sigma \gamma]}+S^{\gamma [\sigma \rho]}+S^{\sigma [\gamma \rho]}) . \label{belsuper}
\end{equation}

\noindent
In electrodynamics this contribution is defined as $S^{\gamma [\alpha\beta]}= \frac{\partial \mathcal{L}}{\partial \partial_\gamma A^\mu} [\eta^{\alpha \mu} A^{\beta}-\eta^{\beta \mu} A^{\alpha}]$. Therefore we have,

\begin{equation}
S^{\gamma [\alpha\beta]}= - F^{\gamma\mu} [\delta^{\alpha}_{ \mu} A^{\beta}-\delta^{\beta}_{ \mu} A^{\alpha}]. \label{spinang}
\end{equation}

\noindent
Inserting (\ref{spinang}) into the Belinfante superpotential (\ref{belsuper}) we have $b^{[\rho\gamma]\sigma} = - F^{\rho\gamma} A^{\sigma}$. But this is the same superpotential found from (\ref{backsuper})! Thus for the Belinfante procedure applied to electrodynamics we have,

\begin{equation}
T^{\mu\nu}_B = T^{\mu\nu}_C + \partial_\alpha (F^{\alpha\mu} A^\nu). \label{belindirect}
\end{equation}

\noindent
This differs from the accepted energy-momentum tensor (\ref{electrodynamicsemt}), according to (\ref{backsuper}), by on-shell terms\footnote{On-shell equivalence of the Belinfante and Hilbert tensors is a well established result \citep{forger2004,pons2011}. The Hilbert tensor in Minkowski spacetime is the accepted energy-momentum tensor in cases such as electrodynamics (but not in general, see \cite{baker2021a}), thus for electrodynamics the Belinfante-Hilbert relationship can be used to obtain the Belinfante superpotential and associated on-shell terms in (\ref{belindirect}) which are required to correctly improve the canonical Noether tensor.}.

Therefore the accepted energy-momentum tensor (\ref{electrodynamicsemt}) is related to the Belinfante tensor (\ref{belindirect}) as follows:

\begin{equation}
T^{\mu\nu} = T^{\mu\nu}_B - A^\nu  E^\mu.
\end{equation}

\noindent The Belinfante prescription alone does not yield the correct expression without adding this additional term ($- A^\nu  E^\mu$) proportional to the equations of motion; equivalence to $T^{\mu\nu}_B$ alone can only be established after imposing the on-shell condition $E^\mu = 0$. Note that requiring such an on-shell condition for just formulating the energy-momentum tensor is a severe restriction; in contrast, the Noether energy-momentum tensor directly obtained in Method II can be defined \textit{without} any on-shell condition --- only conservation requires imposition of the equations of motion.

So we see that, when interested in symmetric energy-momentum tensors, the Belinfante symmetrization procedure does provide an on-shell procedural fix in e.g. the case of electrodynamics. But it is just a symmetrization procedure; it is not clear how it would, for instance, help to obtain the physical (i.e. also tracefree and gauge-invariant and not just symmetric) energy-momentum tensor in general.\footnote{\cite{blaschke2016} provide an improvement procedure based on requiring gauge invariance instead of just symmetry.} Furthermore, it is limited in scope:  it focuses only on the energy-momentum tensor, at the expense of other conserved quantities. By contrast, Method II treats all conserved currents on the same footing; there is no privileging of any specific quantity as most fundamental --- all conserved currents follow directly and uniquely from Noether's first theorem.

If we take the most charitable possible view of Method I, that the ad-hoc improvements are entirely physically justified and a necessary correction after applying Noether's first theorem, one unavoidable fact remains: the improvement procedure still requires on-shell conditions to equate the accepted $T^{\mu\nu}$ to the improved tensor. In Method II, no on-shell conditions are required to obtain the complete set of conservation laws.

\subsection{Method II: Including Gauge Symmetries} \label{SectionMethodII}

There is another method for deriving the energy-momentum tensor, such that we directly obtain it from Noether's first theorem. 

Instead of using the restrictive condition of Method I where we only consider the canonical transformations (\ref{canonicaltrans}), we instead use the most general picture of all possible field transformations such as outlined in (\ref{alltrans}). Considering the transformations of the action 

\begin{equation}
\bar{\delta A_\alpha} = - F^\nu_{\ \alpha} \delta x_\nu, \label{propertrans}
\end{equation}

\noindent
\cite{bessel1921erhaltungssatze} (as well as \cite{eriksen1980,montesinos2006}, among others) derived all 15 accepted conservation laws of electrodynamics directly from Noether's first theorem using (\ref{propertrans}), (\ref{noetherfirstelectrodynamics}) and the sourcefree Lagrangian density $\mathcal{L} = - \frac{1}{4} F_{\mu \nu} F^{\mu \nu}$. In other words, (\ref{propertrans}) are what we call proper transformations. 

\begin{equation}
E^\nu \bar{\delta A_\nu} 
+ \partial_\rho \left( 
 [F^{\rho\nu}  F^\beta_{\ \nu} 
- \frac{1}{4} \eta^{\rho\beta} F_{\mu \nu} F^{\mu \nu}] \delta x_\beta
  \right)
 = 0 \label{noether1electrodynamics}
\end{equation}

\noindent
where $E^\nu$ is the Euler-Lagrange expression. Immediately Noether's first theorem leads to the physical energy-momentum tensor (\ref{electrodynamicsemt}) in square brackets.\footnote{Using (\ref{noether1electrodynamics}) the 15 conformal conservation laws of electrodynamics are immediately obtained by inserting $\delta x_\beta$ from (\ref{conformaltrans}): four from the divergence of the energy-momentum tensor $T^{\rho\beta} = F^{\rho\nu}  F^\beta_{\ \nu} 
- \frac{1}{4} \eta^{\rho\beta} F_{\mu \nu} F^{\mu \nu}$, six from the divergence of the angular momentum tensor $M^{\rho\mu\nu} = x^{\mu}T^{\rho\nu} - x^{\nu}T^{\rho\mu}$, one from the divergence of the dilatation tensor $D^\rho = T^{\rho\beta} x_\beta$ and four from the divergence of the conformal tensor $C^{\rho\alpha} = T^{\rho\beta} (2 x_\beta  x^\alpha - \delta^\alpha_\beta x_\lambda x^\lambda)$.} More precisely, for the case of the 4-parameter Poincar\'e translation, $\delta x_\beta = a_\beta$ is a constant that can be pulled out of the total divergence leaving the elegant identity,

\begin{equation}
E^\nu \bar{\delta A_\nu} 
+ a_\beta \partial_\rho T^{\rho\beta}
 = 0 
\end{equation}

\noindent
Immediately we have a compact identity relating the Euler-Lagrange expression $E^\nu$ and energy-momentum tensor $T^{\rho\beta}$ of electrodynamic theory. The Lorentz force law and Poynting's theorem are compactly derived alongside Maxwell's equations. This compact identity makes it easy to appreciate the celebrated elegance of Noether's theorems.

\subsubsection{Deriving the proper transformations from the Bessel-Hagen method}

How can one obtain the proper transformations (\ref{propertrans}) that lead to the physical conservation laws? The various authors \cite{bessel1921erhaltungssatze,eriksen1980,montesinos2006} that independently came to this conclusion used slightly different rationales, largely to do with requiring gauge invariance of the Noether current or requiring gauge invariance of the transformations themselves. We will follow the Bessel-Hagen approach because he was first to present this result, and took advice from Noether herself on his paper. More explicit application of the Bessel-Hagen method to electrodynamics can be found in \cite{baker2021c}. Starting from the general transformations of fields (\ref{alltrans2}) in the case of electrodynamics we have,

 \begin{equation}
\bar{\delta A_\nu} = - \partial^\beta A_\nu \delta x_\beta -  A_\mu \partial_\nu \delta x^{\mu} + \partial_\nu \phi \label{unsolvedphi}
\end{equation}

\noindent
The question Bessel-Hagen asked is how to derive the parameter $\phi$ such that we have the unique gauge invariant energy-momentum tensor of (sourcefree) electrodynamics. To do this we substitute $\bar{\delta A_\nu}$ into the Noether current (\ref{noetherfirstelectrodynamics}) and solve for $\phi$ to obtain a current which is gauge invariant. The $\phi$ must depend on both the vector potential $A_\alpha$ (in order to obtain a gauge invariant current) and the infinitesimal transformations of coordinates $\delta x^\alpha$ (in order to factor out the 4 parameter Poincaré translation from the current and obtain the energy-momentum tensor). Bessel-Hagen solved for $\phi$, obtaining for the gauge parameter $A_\mu \delta x^\mu$, which is the most trivial scalar combination of the required components. Inserting this $\phi$ into (\ref{unsolvedphi}) and differentiating the third term we have,

\begin{equation}
\bar{\delta A_\nu} = - \delta x_\beta \partial^\beta A_\nu  - A_\mu \partial_\nu \delta x^{\mu} + \delta x_\beta \partial_\nu A^\beta  + A_\mu \partial_\nu \delta x^\mu
\end{equation}

\noindent Remarkably the second and last terms on the right hand side cancel (those associated to the contragredient transformations) and we are left with exactly $\bar{\delta A_\nu} = - F^\beta_{\ \nu} \delta x_\beta$ as in (\ref{propertrans})! Therefore the proper transformations that directly yield the physical conservation laws can be thought of as a mixing of the various symmetries of the action, as opposed to an independent application of symmetries as in the case of the canonical Noether energy-momentum tensor or Noether's second theorem.

We note that the selection of the gauge parameter, while aided by knowledge of the unique gauge invariant energy-momentum tensor in the case of electrodynamics, can be obtained from the Noether current. For this reason the method applies more generally to models where the energy-momentum tensor is not already known. The more general application of Bessel-Hagen to exactly gauge invariant actions in this way is the subject of \cite{baker2021c}, in which the Bessel-Hagen method has successfully been applied to several field theories such as Yang-Mills, Kalb-Ramond, third rank antisymmetric fields, and linearized Gauss-Bonnet gravity.

The proper form of the transformation was noticed for Yang-Mills theory by \cite{jackiw1978}, without deriving this from a procedure such as Method II. While the vast majority of textbooks give the canonical picture alone, some, such as \cite{burgess2002,scheck2012}, have noticed the proper transformations and avoided the restrictive canonical presentation. One of our goals in the following is to settle this ambiguity in favor of the proper transformations through appeal to the converse of Noether's first theorem. 

\subsubsection{Proper transformation as gauge-invariant transformations}

\label{section:righttransformation}

We now know that the appropriate choice of $\delta \bar{A}_{\alpha} = \delta_{C} A_\alpha + \delta_{g} A_\alpha$ in the Noether identity for classical electrodynamics directly leads to the accepted energy-momentum tensor, and that the proper transformation can be chosen by solving for a $\delta_{g} A_\alpha$ that makes the current invariant. We will explore how to justify the specific choice of $\delta_{g} A_\alpha$ a posteriori via the converse of Noether's first theorem in the next section, i.e. by starting from the accepted energy-momentum tensor. 

Before doing so, we want to explore how to motivate the choice of $\delta_{g} A_\alpha$ other than by solving the Noether identity for $\delta_{g} A_\alpha$ while requiring that the energy-momentum is gauge-invariant. To this end, we will consider \citet{eriksen1980}, who argued that gauge invariance of the transformation $\bar{\delta A_\alpha}$ is the property one can use to determine the proper transformation $\bar{\delta A_\alpha} = - F^\nu_{\ \alpha} \delta x_\nu$ as in (\ref{propertrans}).
\citet{eriksen1980} starts from the gauge condition in the case of sourcefree electrodynamics, $\delta_{g} A_\alpha = \partial \chi$ with $\chi = \chi(A)$. The parameter $\chi$ is taken to be an arbitrary gauge parameter we must solve for based on the condition that $\delta_{g} A_\alpha$ must be gauge invariant. By combining the $\delta_T A_\alpha$ (contragredient) and $\delta_g A_\alpha$ (gauge) transformations, the authors find an equation for $\chi$ which does not uniquely determine $\chi$; however they choose the ``simple" solution that leaves $\delta_g A_\alpha$ as a whole gauge invariant, which is identically $\chi = \delta x^\nu A_\nu$, exactly what was found by Bessel-Hagen!

One could now note how intuitive the requirement of a gauge-invariant transformation is: as long as all expressions in the assumptions of the Noether theorem are gauge-invariant, the resulting energy-momentum tensor should come out as gauge-invariant too. However, there exist cases where the proper transformations that are used to derive the unique energy-momentum tensor for a theory are not themselves gauge invariant, as we will discuss in Section \ref{ApplicationsSection}. This indicates limits in the scope of application of \citet{eriksen1980}'s method. We note that the Bessel-Hagen method works more broadly because it treats both the cases where the transformations themselves are gauge invariant, as well as cases where they are not.

\subsection{Converse of Noether's first theorem as a test for Noetherian currents \label{SectionNoetherConverseElectrodynamics}}

We now use the converse of Noether's first theorem relative to the Lagrangian density of electrodynamics and the accepted energy-momentum tensor (\ref{electrodynamicsemt}) in order to arrive at the relevant variational symmetry linked to this $T^{\mu\nu}$. As we will see, the converse can generally be used to decide whether an energy-momentum tensor can be directly derived from Noether's first theorem --- and thus from Method II.

We can derive the form of the transformations $\bar{\delta A_\nu}$ using the converse of Noether's first theorem based on the accepted energy momentum tensor, (\ref{electrodynamicsemt}), as follows: We start with

\begin{equation}
E^\nu \bar{\delta A_\nu} + a_\nu \partial_\mu T^{\mu\nu} = 0 \label{knownidentity}
\end{equation}

\noindent and the Noether identity (\ref{noetherfirstelectrodynamics}). Since the 4-parameter Poincar\'e translation $\delta x_\beta = a_\beta$ is associated to the energy-momentum tensor it follows that $\bar{\delta A_\nu} = U^{\beta}_\nu a_\beta$, namely the transformation of fields must be proportional to the 4-parameter $a_\beta$. Therefore, we must solve for $U^{\beta}_\nu$,

\begin{equation}
E^\nu \bar{\delta A_\nu} 
+ a_\beta \partial_\rho \left( 
- F^{\rho\nu} U^{\beta}_\nu
- \frac{1}{4} \eta^{\rho\beta} F_{\mu\nu} F^{\mu\nu} 
  \right)
 = 0 \label{converseunknown}
\end{equation}

\noindent Subtracting the two equations (\ref{knownidentity}) and (\ref{converseunknown}) we have,
\begin{equation}
 a_\beta \partial_\rho \left( 
- F^{\rho\nu} U^{\beta}_{\ \nu}
- \frac{1}{4} \eta^{\rho\beta} F_{\mu\nu} F^{\mu\nu} 
  \right)
 =  a_\beta \partial_\rho \left(   F^{\rho\nu} F^\beta_{\ \nu} - \frac{1}{4} \eta^{\rho\beta} F_{\mu\nu} F^{\mu\nu} 
  \right)
\end{equation}
\noindent We have $U^{\beta}_\nu = - F^\beta_{\ \nu}$, and thereby recover the proper transformations for the Poincar\'e translation $\delta \bar{A}_\nu = - F^\beta_{\ \nu} a_\beta$. More generally we can solve for $\delta x_\nu$ from the Noether identity and again we have the requirement $\bar{\delta A_\alpha} = - F^\nu_{\ \alpha} \delta x_\nu$. Thus, if we consider the converse of Noether's first theorem on the accepted $T^{\mu\nu}$ (\ref{electrodynamicsemt}), the canonical transformation (\ref{canonicaltrans}) associated to the canonical Noether energy-momentum tensor (\ref{electrodynamicscanonical}) never appears in isolation! In other words, in the case of electrodynamics, the converse of Noether's first theorem supports Method II. At the same time, we can now see that Method I uses the wrong symmetry to begin with. The failure to recognise the properly adapted symmetry transformations leads to the need to introduce --- and justify, if possible --- ad hoc ``improvements''.

The lesson from electrodynamics generalises: Given a proposed energy-momentum tensor, we learn through Noether's converse which (if any) symmetries are linked to it; if there are none linked to it, then the energy-momentum tensor cannot be derived directly from Noether's first theorem.\footnote{This raises the question whether such an object earns the title of energy-momentum tensor in the first place.}

So, in cases like electrodynamics for which the canonical energy momentum tensor lacks essential properties, we find that the improvements can be avoided by using Noether's first theorem properly (that is, by exploiting the complete set of variational symmetries of the action) --- and thus that there is nothing wrong with the Noether method to begin with. In cases where the converse does not give symmetries linked to an energy-momentum tensor, we at least learn that this energy-momentum tensor cannot be derived from Noether's first theorem.

We acknowledge that the converse approach we are advocating is not as straightforward if we do not already know the appropriate energy-momentum tensor for the relevant fields.  We have criticized the improvement approach in Method I because it relies on knowing the proper form of the energy-momentum tensor in order to find the appropriate improvement terms; yet, we also need to know the proper form of the energy-momentum tensor to obtain the proper variational symmetries from the converse of Noether's first theorem.  If the proper energy-momentum tensor is not uniquely known (several candidates exist), this can lead to a kind of reflective equilibrium in assessing candidates for energy-momentum tensors and the associated symmetries. We will turn to just such a case in the next section, namely linearized spin-2 fields where numerous different energy-momentum tensors have been proposed.

\section{Beyond electrodynamics}
 \label{ApplicationsSection}

Up to this point we have used electrodynamics to explicate Method II --- but it has much broader scope. A recent series of papers has shown how Method II applies to several classical, relativistic field theories, such as:
\begin{itemize}
	\item \emph{(Source-free) Yang Mills} \citep{baker2021c}, with the Lagrangian $\mathcal{L}_{YM}=-\frac{1}{4}F_{\mu\nu}^{a}F_{a}^{\mu\nu}$.  Applying Method II with the ``mixed" variational symmetry $\bar{\delta}A_{\mu}^{a}=-F_{\mu\nu}^{a} \delta x^{\nu}$ (with $\delta x^{\nu} = a^{\nu}$) leads to the energy-momentum tensor $T^{\mu\nu} = F_a^{\mu\lambda}F_{\ \ \ \lambda}^{a \nu}-\frac{1}{4}\eta^{\mu\nu}F_{\lambda\rho}^{a}F_{a}^{\lambda\rho}$.\footnote{The field strength tensor is given by $F_{a\mu\nu}=\partial_{\mu}A_{av}-\partial_{\nu}A_{a\mu}+C_{abc}A_{\mu}^{b}A_{\nu}^{c}$ where $C_{abc}$ is the totally antisymmetric structure constant.} The energy-momentum tensor is invariant under the gauge transformation $\delta_{g}\bar{A}_{a\mu}=\partial_{\mu}\theta_{a}+C_{abc}^{~}A_{\mu}^{b} \theta^{c}$.
	\item \emph{Linearized Gauss-Bonnet gravity} \citep{baker2019,baker2021b}, with the Lagrangian $\mathcal{L} = \frac{1}{4} (R_{\mu\nu\alpha\beta} R^{\mu\nu\alpha\beta} - 4 R_{\mu\nu} R^{\mu\nu} + R^2)$.\footnote{Appearing in the Lagrangian are the linearized Riemannian tensor, defined as $R^{\mu\nu\alpha\beta} = \frac{1}{2} (  \partial^\mu \partial^\beta h^{\nu\alpha} + \partial^\nu \partial^\alpha h^{\mu\beta} -\partial^\mu \partial^\alpha h^{\nu\beta} - \partial^\nu \partial^\beta h^{\mu\alpha})$, and contractions of it.} The variational symmetry $\delta \bar{h}_{\rho\sigma} = - 2 \Gamma^\nu_{\ \rho \sigma} \delta x_\nu$ (with $\delta x^{\nu} = a^{\nu}$) leads to the generally accepted energy-momentum tensor,\footnote{Explicitly, $T^{\omega\nu} = -  R^{\omega\rho\lambda\sigma} R^\nu_{\ \rho\lambda\sigma} + 2  R_{\rho\sigma} R^{\omega \rho \nu \sigma} + 2  R^{\omega\lambda} R^\nu_{\ \lambda} -  R R^{\nu\omega} + \frac{1}{4} \eta^{\omega\nu}(R_{\mu\lambda\alpha\beta} R^{\mu\lambda\alpha\beta} - 4 R_{\mu\nu} R^{\mu\nu} + R^2)$}  which is gauge-invariant under the spin-2 gauge transformation $\delta_g \bar{h}_{\mu\nu} = \partial_\mu \xi_\nu + \partial_\nu \xi_\mu$.
\end{itemize}  
Just as with electrodynamics, applying the mixed variational symmetry in each of these cases leads directly to the accepted energy-momentum tensor. \citet{baker2021c} discusses several other cases as well.

What can then be said about the scope of the method? For a given gauge invariant Lagrangian density, an exact variational symmetry can be found such that the Noether current associated to the Poincar\'e translation will be the physical energy-momentum tensor. Notably, it is \textit{not} a necessary criterion that the total symmetry transformation is gauge-invariant: Recalling \ref{section:righttransformation}, the decisive symmetry transformation in electrodynamics (as given by the standard Lagrangian) is gauge-invariant, but the symmetry transformation is not always itself gauge invariant (e.g. linearized Gauss-Bonnet gravity). This means that the proper variational symmetries can not always be systematically obtained from requiring gauge-invariance of the proper transformation, which showcases the restrictions of the procedure presented in section \ref{section:righttransformation}. In other words, an exactly gauge invariant symmetry transformation will only be sufficient but not necessary for obtaining a gauge-invariant conserved current; Method II à la Bessel-Hagen has a much wider scope than the method of \cite{eriksen1980}.

Put the other way around, problems arise for Method II when: (1) we have a model that does not have an exactly gauge invariant action, so that solving for the right gauge-transformation becomes problematic, or (2) when the energy-momentum tensor is questionable but the BH method is not applicable because there is just no gauge symmetry to begin with. We will outline now how in both cases at least the Noetherian\footnote{See \S \ref{section:Equivalence} for discussion of Noetherian vs. non-Noetherian currents.} nature of the energy-momentum tensor can be checked upon application of Noether's converse.

With respect to (2), an interesting application of Noether's converse is to reveal the improved Callan-Coleman-Jackiw (CCJ) traceless energy-momentum tensor of the Klein-Gordon theory (see \cite{callan1970}) as non-Noetherian relative to the standard Lagrangian.\footnote{We call an energy-momentum tensor Noetherian relative to $\mathcal{L}$ if it is directly derivable from a variational symmetry of $\mathcal{L}$ via Noether's identity.}

However, an energy-momentum tensor that is non-Noetherian relative to some Lagrangian may be Noetherian to the same Lagrangian up to a divergence term; this is exactly the case of the CCJ-energy-momentum tensor (see \cite{kuzmin2001}).

Turning to (1), Noether's (original) theorem involves the invariant action condition $S = S'$, namely that an action does not change after simultaneous transformation of the independent variables (e.g. coordinates) and dependent variables (e.g. fields). This condition requires strict invariance, namely the exact symmetries which we have referred to throughout the article. In this case Noether's theorem can be applied as in her original paper, in a straightforward manner which we consider for both Method I and Method II. There are generalizations of this theorem to transformations which do not satisfy this action condition, and instead after simultaneous transformation of the independent and dependent variables differ by a boundary term, e.g. $S' - S = \int \partial_\mu B^\mu dV$. Therefore in cases where an action has an inexact symmetry (transforming the action results in a boundary term), we cannot trivially apply Noether's first theorem as in her article, and must turn to one of the generalizations of invariance up to a boundary term. These ``modern'' generalizations originate with the other result in Bessel-Hagen's paper, invariance up to boundary terms, but extend to the so-called MAO theorem, which we will discuss in \S \ref{section:Equivalence}. We do not apply these generalizations in our article because our main concern is to compare the conventional application of Noether's first theorem (Method I) to the lesser known method (Method II), both in the context of the action condition $S = S'$. However, we will now discuss a possible valuable application of these generalizations, which we could not find any studies of in the literature.

An appealing possible application with respect to (1) is the case of the spin-2 Fierz-Pauli action where the gauge symmetry of the equation of motion (linearized diffeomorphisms) is not an exact symmetry of the action; the action is only invariant up to a boundary term (see \cite{baker2019}). Furthermore, there is no gauge-invariant energy-momentum tensor for spin-2 gravity (see \cite{magnano2002symmetry}) to begin with, so we cannot use gauge invariance to help pick out a unique expression.\footnote{Arguably, this jeopardises the application of BH method which is centrally about achieving a gauge-invariant current.} If an action is not exactly invariant such as in the case of spin-2, generalizations of Noether's theorem to symmetries up to boundary terms (i.e. the inexact symmetries method in \cite{bessel1921erhaltungssatze}) must be applied; the application of these methods to spin-2 Fierz-Pauli theory is the subject of future work.

To elaborate a bit on the issue: For linearized gravity (massless spin-2 gravity), there are numerous proposals for $T^{\mu\nu}$ (see \cite{bicak2016} for an overview). This ambiguity cannot be avoided in, for example, attempts to derive general relativity from a spin-2 field theory that proceed by taking the spin-2 field $h_{\mu\nu}$ to be self-coupled. Which $T^{\mu\nu}$ should be added to the action to represent this self-coupling?  Here authors disagree on whether the Einstein field equations can be derived from spin-2 Fierz-Pauli theory, to a large degree based on their choice of which $T^{\mu\nu}$ to select (if they even grant that it is physically well-defined despite its inevitable gauge-dependent nature). (See \cite{padmanabhan2008} for a criticism, and \cite{barcelo2014}  for a defense, of conventional wisdom on this issue.)

Linearized (massless) spin-2 gravity is given by the (massless) Fierz-Pauli Lagrangian density 

\[\mathcal{L}_{FP} = \frac{1}{4} \left( \partial_{\alpha} h^{\beta}_{\beta} \partial^{\alpha} h^{\gamma}_{\gamma} - \partial_{\alpha} h_{\beta \gamma} \partial^{\alpha} h^{\beta \gamma} + 2 \partial_{\alpha} h_{\beta \gamma} \partial^{\gamma} h^{\beta \alpha} - 2 \partial^{\alpha} h^{\beta}_{\beta} \partial^{\gamma} h_{\gamma \alpha} \right)\]

\noindent
with canonical Noether energy-momentum tensor $T^{\mu\nu}_C = \eta^{\mu\nu} \mathcal{L} - \frac{\partial \mathcal{L}}{\partial (\partial_\mu h_{\alpha\beta})} \partial^\nu h_{\alpha\beta}$ --- the linearized Einstein energy-momentum tensor gives exactly this canonical expression (\cite{szabados1992canonical}). Strikingly, this tensor is neither gauge-invariant nor symmetric nor traceless. As we know from \cite{magnano2002symmetry}, there is no energy-momentum tensor for spin-2 Fierz-Pauli theory that is gauge-invariant. Candidates for such improved energy-momentum tensors for linearized gravity usually presented include, for example, the linearized Hilbert and Landau-Lifshitz expressions --- both of which can be obtained by adding the appropriate divergence of superpotential and terms proportional to the equations of motion to $T^{\mu\nu}_C$ (see \cite{baker2021cqg}). Strong adherents to Method I might then suggest that since any such energy-momentum tensor (conserved on-shell using the spin-2 Fierz-Pauli equation of motion) follows from the addition of improvement terms, that all such linearized gravity energy-momentum tensors are in some sense connected to Noether's first theorem. \cite{baker2021cqg} shows that there are infinitely many such improved energy-momentum tensors for linearized gravity.

A much more straightforward approach for spin-2 linearized gravity then is to apply the converse of Noether's first theorem to the various expressions in the literature as we did for electrodynamics in the previous section. This would concretely determine which (if any) can yield $\delta X_\beta$ and $\bar{\delta h_{\mu\nu}}$ symmetry transformations to prove a direct and meaningful connection to Noether's first theorem. Concretely, one would have to use the Noether identity

\begin{equation}
\left( \frac{\partial \mathcal{L}}{\partial h_{\mu\nu}}
- \partial_\rho \frac{\partial \mathcal{L}}{\partial (\partial_\rho h_{\mu\nu})} \right) {\delta \bar{h}_{\mu\nu}} 
+ \partial_\rho \left( 
\eta^{\rho\beta} \mathcal{L} \delta X_\beta
+ \frac{\partial \mathcal{L}}{\partial (\partial_\rho h_{\mu\nu})}{\delta \bar{h}_{\mu\nu}} 
  \right)
 = 0 \label{Noether1}
\end{equation}

\noindent
for each energy-momentum tensor and solve for $\delta \bar{h}_{\mu\nu}$ in the same way we solved for the variational symmetries of electrodynamics in Section \ref{SectionNoetherConverseElectrodynamics}, where $\bar{\delta h_{\mu\nu}} = - \partial^\beta h_{\mu\nu} \delta X_\beta + \delta h_{\mu\nu}$. As in the case of electrodynamics, the term proportional to the Lagrangian density ($\eta^{\rho\beta} \mathcal{L}$) must have a Lagrangian density $\mathcal{L}$ which yields the spin-2 equation of motion in the Euler-Lagrange equation --- otherwise the energy-momentum tensor in question will not be associated to spin-2 Fierz-Pauli theory in the context of Noether's first theorem, regardless of the transformations we consider. Regardless of the outcome of this calculation we will have a strong statement about the energy-momentum tensors for spin-2 theory in the literature: either that Noether transformations can select a preferred expression, confirm numerous expressions can be obtained from the Noether approach, or show that there are problems with applying Noether's first theorem to this model as a whole. If transformations can uniquely be solved for by the converse of Noether's first theorem, then we can say that a given expression can be directly derived. If there is no solution, then a given expression cannot be claimed to be associated to Noether's first theorem for a particular Lagrangian density. Whether or not the various published expressions can be directly obtained, regardless of outcome, will provide clear insight into the relationship between the linearized gravity energy-momentum tensors in the literature and Noether's first theorem. The treatment of energy in linearized gravity may have some bearing on active disputes regarding energetic quantities in GR (see, for instance, \cite{read2020functional,duerr2019against}). This application of the converse of Noether's first theorem to spin-2 is the subject of future work.\niels{reviewer: While I appreciate the content of this section, I worry that it upsets the narrative flow of the argument. Especially given that end of section 4 flows into section 6 quite naturally NL: we could move this to after section 6?}

\section{Equivalence classes}

\label{section:Equivalence}

Mathematics often draws finer distinctions than physics requires. Physicists typically treat a unique definition of a fundamental quantity as necessary for understanding its physical significance. This can conflict with the embarrassment of riches resulting from new frameworks produced by mathematicians, particularly when they lead to quantities with an ambiguous physical status.\niels{Harvey's point 3: 3. On p. 22 we read: ``Recent philosophical discussions (see Brown (2020) in this volume) have argued that Noether’s theorem should be read as relating an equivalence class of symmetries to an equivalence class of conservation laws.” First, a somewhat pedantic point. Brown does not strictly do this; he summarises the independent work of Martines Alonso (1979) and Olver (1986) whose theorem is not equivalent to Noether’s theorem I nor a re-reading of it. (I am also unaware of other recent philosophical discussions of what Brown calls the MAO theorem.) More importantly, the authors argue that the motivation for introducing equivalence classes in this context results from “mathematicians’ drive to generalise”. I think this claim is questionable. I don’t know what the motivations of Martines Alonso and Olver were, but the result was a Noether-like theorem which achieved something not found in Noether’s machinery, a one-to-one correspondence between variational symmetries and conservation laws relative to a given Lagrangian— the price being paid being that the correspondence is actually between suitably defined equivalence classes of these principles. NL: fair enough. Brown (continued): The authors put heavy emphasis on Noether’s own (first) converse theorem in their section 4.3, which might appear to make the MAO result redundant. But this converse theorem does not connect conservation principles with variational symmetries, but rather it connects a given form of the Noether current with symmetries. It is, as the authors are fully aware, an off-shell condition. The significance of the MAO result as I see it is that by introducing “equivalence” classes the on-shell version of the converse theorem, involving genuine conservation laws, is obtained. I note in this connection that on p. 7. line 3 in section 3, I think it is better to call $E^A$ the Euler-Lagrange expression (Noether called it the Lagrange expression). (See also in this connection pp. 9, 11, 13, 15, 16.) The “Euler-Lagrange equation” $E^A = 0$ does not feature in equation (2). So it is misleading to state that the converse of Noether’s theorem establishes a relationship that “holds off-shell between conserved currents and variational symmetries” (p. 24 penultimate paragraph): why conserved? NL: I have removed the ``conserved".} 
A natural response is to regard some range of mathematically distinguished possibilities as falling within an equivalence class, such that physical interpretations need not draw distinctions among its members. Recent philosophical discussions by \cite{brown2020} (in this volume) have pointed out that modern Noether-\textit{like} theorems by Mart\'{i}nez, Alonso and Olver (dubbed the MAO theorem by Brown) should be read as relating an equivalence class of symmetries to an equivalence class of conservation laws.\footnote{To be more precise, \cite{olver2000applications}'s statement of Noether's theorem (Theorem 5.58, p. 334) introduces the following notion of equivalence between variational symmetries --- transformations on the space of dependent and independent variables of the differential equations, in the form of a (generalized) vector field, that preserve the action. Two variational symmetries are equivalent if they differ by a trivial symmetry; a trivial symmetry is a transformation on the space of solutions that is just the generator of the dynamical flow (up to parameterization). The generalized notion of a vector field includes ``evolutionary vector fields''; a trivial symmetry transformation corresponds to an evolutionary vector field that vanishes on the solutions of the differential equation under consideration.  The need to ``quotient out'' by trivial symmetries arises as a result of the shift to a different mathematical setting.} Perhaps we could follow a similar strategy and elucidate only an equivalence class of energy-momentum tensors, setting aside the search for a unique energy-momentum tensor we have been pursuing as misguided or unnecessary? In this section we aim to adjudicate these questions regarding uniqueness and the appropriate criteria of equivalence, or (perhaps more accurately) at least to survey some of the considerations that bear on them.

To make this concrete, suppose we treat two candidate energy-momentum tensors as equivalent iff they differ by superpotential terms and linear combinations of the equations of motion. In the case we have focused on in Method II, this equivalence class consists of tensors generated from the standard energy-momentum tensor (\ref{electrodynamicsemt}) by adding the divergence of a superpotential and terms proportional to the equations of motion:\footnote{Conventionally these ``improvements'' (the divergence of a superpotential and terms proportional to the equations of motion) are applied to the canonical Noether tensor (\ref{electrodynamicscanonical}) as in (\ref{backsuper}) for Method I, but they can be applied to any Noether current (such as in (\ref{allimprovements}) using Method II), satisfying the Noether identity (\ref{noetherfirstelectrodynamics}) on-shell.}
\begin{equation}
T_{G}^{\mu\nu} = F^{\mu\alpha} F^\nu_{\ \alpha} - \frac{1}{4} \eta^{\mu\nu} F_{\alpha\beta} F^{\alpha\beta} + C_1 \partial_\alpha \Psi^{[\mu\alpha]\nu} + C_2 A^\nu  E^\mu + C_3 A^\mu  E^\nu + C_4 \eta^{\mu\nu} A_\alpha E^\alpha \label{allimprovements}
\end{equation}
\noindent where $\Psi^{[\rho\alpha]\sigma}$ is the most general rank-three tensor defined in terms of the potential and derivatives of the potential, with the required symmetry properties. The subscript indicates that this is the general form for tensors in this equivalence class, with the coefficients $C_n$ taking arbitrary values. We trivially recover the standard expression (\ref{electrodynamicsemt}) by setting $C_n=0$ (for all $n$) because Method II yields (\ref{electrodynamicsemt}) without requiring the $C_n$ terms.  But even allowing $C_n$ to take arbitrary values, (\ref{allimprovements}) will still satisfy the Noether identity (\ref{noetherfirstelectrodynamics}) on-shell.\footnote{In the case of spin-2 Fierz-Pauli theory, any of the published energy-momentum tensors for linearized gravity can be obtained from the canonical Noether tensor by adding superpotential terms and on-shell contributions \citep{baker2021cqg}; there are infinitely many energy-momentum tensors in this equivalence class.} Therefore despite the mathematical freedom to allow for an on-shell equivalence class, we see that the equivalence class is not at all needed to yield the unique $T^{\mu\nu}$ of the theory.

The point generalizes to other conserved currents.  Based on our analysis above, we see more generally that Noether's theorem and its converse specify a relationship between a properly ``Noetherian'' current $J$ and the variational symmetries of the action that hold off-shell, regardless of whether the Euler-Lagrange equations are satisfied.  But just as we can add the terms with coefficients $\{C_1, ... , C_4\}$ to $T_G$ above, we can add terms to other Noether currents:  schematically, $J_G = J + J_{1} + J_{2}$, for currents $J_{1}$ that can be expressed as the divergence of a superpotential and terms $J_{2}$ proportional to the equations of motion. The divergence of the $J_{1}$ terms vanish in the Noether identity by construction, regardless of the features of the Lagrangian; the $J_2$ terms vanish on-shell. We classify these further terms as ``non-Noetherian'' because applying the converse of Noether's theorem does not yield any relationship to variational symmetries of the Lagrangian. The proposed definition of equivalence then amounts to generating an equivalence class of conserved currents by adding ``non-Noetherian'' terms.

There are plausible empiricist motivations in favour of this definition of equivalence.  We only have empirical access to quantities defined on-shell:  all empirical data is in the solutions, so to speak.  Alternatively, this data can be seen as codified into the dynamical equations and the space of possible initial conditions (provided the problem is well-posed).  Since the collection $\{T_G\}$ (or $\{J_G\}$) and the Noetherian currents agree on-shell, by construction, we cannot measure the differences among them, and it makes sense to treat them as belonging to an equivalence class.

But \textit{should} we indeed treat the symmetries as determining only an equivalence class of energy-momentum tensors, or currents, in this empiricist sense? A positive answer would contrast sharply with the common practice in physics of taking (\ref{electrodynamicsemt}) as the ``correct'' expression for the energy-momentum tensor, and of delineating properly Noetherian currents. One possibility is that physicists simply choose one element of the equivalence class by convention or as a matter of convenience. (If the elements of the equivalence class truly ``represent the same physical situation,'' it would be a mistake to demand physical justification of the choice.) But this is not the position one finds; instead, there are active debates regarding, for example, what is the uniquely correct physical expression for the energy-momentum tensor for various classical field theories \citep{gotay1992,forger2004}.

There seems to be a more insightful explanation than that of a mere conventionalist, or of (unfounded) convenience, as to why practitioners do not adopt such empiricist equivalence classes. To see this, we first note that this empiricist definition of equivalence builds on a sort of provincialism on what to count as empirically relevant.  This can be challenged when we consider a given classical field theory in a somewhat broader theoretical context, such as how it relates to other theories. While off-shell differences may not be empirically relevant classically, they do for instance become empirically relevant upon quantisation (think of the Feynman path integral picture). Furthermore, regarding the energy-momentum tensor in particular, coupling to gravity does render the absolute value of matter energy-momentum empirically meaningful, suggesting that we should care about the absolute value even outside of gravitational theories (in other words, we should not be indifferent towards shifts in superpotential either).  

But we need not turn to the relationships to other theories for reasons to reconsider what should qualify as an empirically meaningful quantity, as \cite{Rovelli}'s take on gauge symmetry illustrates.  Conventional wisdom treats gauge-dependent quantities as redundant, ``descriptive fluff''; the collection of states related by gauge transformations are treated as elements of an equivalence class. We should then characterize the state of a system entirely in terms of gauge-invariant observables, even if gauge-dependent quantities appear as a useful bookkeeping device in some calculations. \cite{Rovelli} argues that this view betrays a limited conception of how the theory represents systems.  Briefly put, it makes sense if we take the theory to represent isolated systems, but when we use gauge theory to describe interacting subsystems the gauge degrees of freedom are essential to describing coupling among these systems.  Without going further into the details, we hope that the main point relevant to our discussion is clear: interpretative questions, e.g. whether the theory represents isolated systems or coupled systems, have to be settled prior to deciding what quantities count as empirically meaningful.  In conjunction with our point in the previous paragraph, namely that relations among theories may also require more fine-grained distinctions, this tells against accepting an equivalence class of energy-momentum tensors. 

Both considerations support a general methodological point of view, namely a preference for individuating physical quantities in terms of deep principles of the theory whenever that is possible. Doing so makes it more likely that the physical quantities will be well-defined in a broader range of contexts, rather than being applicable only to the case at hand, and will continue to be useful as we modify or extend current theory.  We can justify a strong preference for a definition of Noetherian currents, and the preferred form of the energy-momentum tensor, from this point of view as follows.  The expressions  $\{J_G\}, \{T_G\}$ are obtained by adding on ``non-Noetherian'' terms that lack any clear connection with the Lagrangian, and only hold on-shell.  By contrast, Noether's first theorem and its converse, as described in detail above, establish a direct relationship that holds off-shell between currents and the variational symmetries of the Lagrangian. The considerations above support, in line with this methodological principle, a much more stringent criterion of equivalence than the empiricist endorses, and supports the preference for a unique expression for the energy-momentum tensor and other conserved currents.

Before closing this section, we should acknowledge that these interpretative questions depend sensitively on the formal context in which they are pursued, as well as on the general methodological issues we have just highlighted. Above we have used the Lagrangian formulation of classical field theories, such that solutions to the relevant field equations are found via variational methods, and focused on the relationship between variational symmetries and conserved quantities. Noether's theorem can be applied in a variety of formalisms, including to Hamiltonian systems \citep[see][]{butterfield2006symmetry} and directly to the solutions of differential equations (whether or not they have been derived using variational techniques). For the latter case, there are several distinct notions of symmetry, defined in terms of transformations on the space of solutions of the differential equations whose infinitesimal generators have specific properties \cite[e.g.,][]{belot2013}.  (One can think of the differential equation as introducing a structure that these symmetries preserve, where this is more general than a Lagrangian.)  The subtle relationship between these different contexts in which we can discuss symmetries and conservation laws is nicely illustrated by \cite{Smith2008}. In classical particle mechanics, there are several examples of non-standard Lagrangians that are dynamically equivalent to a ``standard'' Lagrangian, of the form ``$L = T - V$," yet with a different variational symmetry. Noether's theorem then yields different conservation laws for the two Lagrangians, even though they yield the same equations of motion. What grounds would there be for choosing among the class of Lagrangians $\{\mathcal{L}_1, \mathcal{L}_2, ...\}$ in order to determine the ``correct'' conserved quantity?  Whether one admits physical grounds to choose among the candidate Lagrangians, and so recover a clear connection between symmetry and conserved quantities, depends on whether we can go above and beyond the dynamical equations themselves.\niels{Harvey's second point: 2. The end of section 6 mentions the case of inequivalent Lagrangians. It would be interesting to know what the authors would make of Sudbery’s 1986 choice of a non-standard, non-Lorentz scalar free electrodynamical Lagrangian in which the standard conserved energy momentum tensor is associated not with traditional translational invariance but with invariance under (internal) duality rotations. If I have understood this result correctly, it does not appeal to gauge invariance! (See A Sudbery, J. Phys. A: Math Gen (1986) L33-L36.)}

\section{Conclusions}
\label{section:Conclusions}

Noether's first theorem is one of the most celebrated results in physics. Yet, standard textbook and literature presentation gives the picture that this method fails to derive standard physical conservation laws: the canonical Noether energy-momentum tensor, which is derived using a restricted condition placed on Noether's first theorem, does not give the known physical energy-momentum tensor in foundational models such as electrodynamics and Yang-Mills theory. All of this creates the impression that Noether's first theorem, despite frequent praise in the scientific community, is in some sense not working in practice for our most significant theories. 
We hope that our presentation of the Bessel-Hagen method (Method II) has let the reader regain confidence in the power of Noether's first theorem when applied to exactly gauge-invariant field theories: using the complete set of (mixed) symmetries of the action (both gauge and coordinate symmetries), one obtains transformations that directly yield the known physical energy-momentum tensor of electrodynamics and theories with a gauge-invariant Lagrangian density more generally. No ``improvement'' of the energy-momentum tensor is needed to supplement (nor actually advised for by) the Noether machinery. 

In showcasing the proper application of Noether's theorem in the context of exactly gauge-invariant theories, we have, moreover, learned that the conventional wisdom that a specific variational symmetry (namely the canonical variational symmetry) is linked to a specific conservation law by the Noether machinery after all remains true within the bulk of classical field theory in practice. As we had already said in the introduction, one is free to question the linkage by more theoretical counterexamples --- but this is a question for another day. Yet another interesting insight was gained along the way: contra common characterisations in the literature, Noether's first theorem is not solely concerned with what we called canonical variational symmetries exclusively but rather the complete set of symmetries of the action (this includes gauge symmetries which are sometimes portrayed as being only associated to Noether's second theorem). We have used the converse of Noether's first theorem as a method for emphasizing this fact, as the canonical variational symmetries do not follow from the converse theorem for the majority of accepted physical energy-momentum tensors in the literature.

Finally, there is a sense in which our overall message in favour of Method II could be made even more strongly: Throughout the article, we had tacitly accepted the common theme in the literature to pay special attention to the energy-momentum tensor over and above other conserved currents in special relativistic field theory. This is important to note as it is quite possible that many of the non-uniqueness and ambiguity problems associated to tensor conservation laws are a result of limiting oneself to the case of energy-momentum tensor specifically, and that the large variety of methods for energy-momentum construction compared to the other tensors is rather an issue that may not be solved by treating the energy-momentum tensor as a privileged standalone object. From the point of view of Noether's first theorem and Method II, none of the standard conserved tensors (energy-momentum, angular momentum, conformal and dilatation) are privileged compared to each other. Thus, if we agreed to consider only methodology which links all of the conserved tensors of a theory to variational symmetries simultaneously, Noether's first theorem in the sense of Method II may give a much needed uniqueness in methodology akin to the Euler-Lagrange equation for an equation of motion. Such a view, if adopted, has promise to end the various ambiguity and non-uniqueness problems associated to the energy-momentum tensor once and for all.

\section{Acknowledgments}

We are grateful to James Read, Bryan Roberts and Nicholas Teh for the invitation to contribute to this volume. We thank Harvey Brown for encouraging comments and helpful suggestions on the paper. We also thank an anonymous reviewer for their helpful input.

\bibliography{NoetherReferences}

\end{document}